\documentclass[a4paper,11pt]{article}

\usepackage[english]{babel}

\usepackage[a4paper,top=2cm,bottom=2cm,left=2.5cm,right=2.5cm,marginparwidth=1.75cm]{geometry}

\usepackage{times}

\usepackage{graphicx}
\usepackage{amsmath}  
\usepackage{mathtools}
\usepackage{booktabs} 
\usepackage{multirow}
\usepackage{authblk}
\usepackage[colorlinks=true, allcolors=blue]{hyperref}

\newcommand{\EE}{e^+e^-}
\newcommand{\MM}{\mu^+\mu^-}

\newcommand{\bcntr}{\begin{center}}
\newcommand{\ecntr}{\end{center}}
\newcommand{\beq}{\begin{equation}}
\newcommand{\eeq}{\end{equation}}
\newcommand{\beqar}{\begin{eqnarray}}
\newcommand{\eeqar}{\end{eqnarray}}
\newcommand{\bitm}{\begin{itemize}}
\newcommand{\benu}{\begin{enumerate}}

\newcommand{\bitmb}{\begin{itemize}}
\newcommand{\benub}{\begin{enumerate}}
\newcommand{\eitm}{\end{itemize}}

\newcommand{\bfrm}{\begin{frame}}
\newcommand{\efrm}{\end{frame}}
\newcommand{\bct}{\begin{center}}
\newcommand{\ect}{\end{center}}
\newcommand{\bclm}{\begin{columns}}
\newcommand{\eclm}{\end{columns}}
\newcommand{\bpic}{\begin{overpic}}
\newcommand{\epic}{\end{overpic}}
\newcommand{\bblk}{\begin{block}}
\newcommand{\eblk}{\end{block}}

\newcommand{\eenu}{\end{enumerate}}

\newcommand{\ohm}{\si{\ohm}}

\newcommand{\gev}{\hbox{GeV}}

\newcommand{\abs}[1]{\left|#1\right|}

\title{\textbf{\fontsize{23pt}{20pt}\selectfont Search for Charged Lepton Flavor Violation at BESIII}\\(Presented at the 32nd International Symposium on Lepton Photon Interactions at High Energies Madison, Wisconsin, USA, Augues 25-29, 2025)}

\author[1]{MingKuan Yuan}
\author[2]{TianZi Song}
\author[2]{ZhengYun You}

\affil[ ]{\vspace{0.5em}\normalsize\bfseries on behalf of the BESIII Collaboration\vspace{0.5em}}

\affil[1]{Institute of Modern Physics, Fudan University, Shanghai, 200433, China}

\affil[2]{School of Physics, Sun Yat-sen University, Guangzhou, 510275, China}

\begin{document}

\maketitle

\begin{abstract}
    Charged lepton flavor violation (CLFV) is forbidden in the Standard Model but predicted by many new physics models. We present searches for CLFV in charmonium decays using world-leading datasets collected by the BESIII detector. The processes $J/\psi\to e\tau$, $J/\psi\to e\mu$, and $\psi(3686)\to e\mu$ are investigated using world-leading $J/\psi$ and $\psi(3686)$ dataset collected by BESIII. No significant signals are observed, and upper limits on branching fractions are set at $\mathcal{B}(J/\psi\to e\tau)<7.5\times10^{-8}$, $\mathcal{B}(J/\psi\to e\mu)<4.5\times10^{-9}$, and $\mathcal{B}(\psi(3686)\to e\mu)<1.4\times10^{-8}$ at 90\% confidence level. These results provide constraints on Wilson coefficients in effective field theory and probe new physics at high energy scales.
\end{abstract}

\section{Introduction}

In the Standard Model (SM), quarks change flavor through electroweak transitions, with rates governed by the Cabibbo-Kobayashi-Maskawa matrix. Similarly, neutrinos oscillate between electron, muon, and tau lepton flavors according to rates given by the Pontecorvo-Maki-Nakagawa-Sakata matrix~\cite{ref::neutrino_1}, a phenomenon beyond the SM. However, lepton flavor violation has not been observed in the charged lepton sector. While non-zero neutrino masses and neutrino oscillations allow charged lepton flavor violation (CLFV) to occur through SM loops, the predicted branching fractions (BFs) are on the order of $\sim10^{-54}$~\cite{Calibbi:2017uvl}, rendering these processes experimentally inaccessible. Therefore, any observation of CLFV would constitute an unambiguous signal of new physics (NP) beyond the SM~\cite{deGouvea:2013zba,Heeck:2016xwg,Beacham:2019nyx}.

The BESIII detector~\cite{Ablikim:2009aa} records symmetric $e^+e^-$ collisions provided by the BEPCII storage ring~\cite{Yu:IPAC2016-TUYA01} in the center-of-mass (c.m.) energy range from 1.84 to 4.95~GeV, with a peak luminosity of $1.1 \times 10^{33}~\text{cm}^{-2}\text{s}^{-1}$ achieved at $\sqrt{s} = 3.773~\text{GeV}$. BESIII has accumulated datasets with world-leading statistics, including 10 billion $J/\psi$ events and 2.7 billion $\psi(3686)$ events~\cite{Liao:2025lth}. These extensive and unique datasets provide an excellent opportunity to probe NP through highly sensitive searches for CLFV decays. In this paper, we present searches for the CLFV decays $J/\psi\to e\tau$, $J/\psi\to e\mu$, and $\psi(3686)\to e\mu$~\cite{BESIII:jpsi2etau,BESIII:jpsi2emu,BESIII:psip2emu}.

\section{Charged lepton flavor violation decay measurements at BESIII}

\subsection{Search for CLFV decay $J/\psi\to e\tau$}

The search for the CLFV decay $J/\psi\to e\tau$ is based on $(10.087\pm0.044)\times10^9$ $J/\psi$ events collected in 2009, 2012, 2018, and 2019 at BESIII. The $J/\psi$ events in 2009 and 2012 is denoted as "data sample I", while the data sample collected in 2018 and 2019 is denoted as "data sample II". The tau lepton in the signal decay is reconstructed through the decay $\tau^+\to\pi^+\pi^0\nu$, where the $\pi^0$ is reconstructed via $\pi^0\to\gamma\gamma$. After applying all selection criteria, the detection efficiency for sample I (II) is determined to be $(20.24\pm0.05)\%$ ($(19.37\pm0.02)\%$).

The dominant background contaminations arise from continuum processes and hadronic $J/\psi$ decays such as $J/\psi\to \pi^+\pi^-\pi^0$. The continuum background is studied using a $150~{\rm pb}^{-1}$ data sample collected at $\sqrt{s}=3.08~\gev$ and a $2.93~{\rm fb}^{-1}$ data sample collected at $\sqrt{s}=3.773~\gev$. The continuum background events are estimated to be $5.8\pm1.8$ ($37.9\pm11.5$) for data sample I (II). The $J/\psi$ decay background is studied using inclusive Monte Carlo (MC) samples, where only a few events survive all selection criteria. Main background processes from $J/\psi\to\pi^+\pi^-\pi^0$, $J/\psi\to\rho\pi$, $J/\psi\to \omega f_2(1270)$, and $J/\psi\to\bar{p}n\pi^+$ are studied with exclusive MC samples. The normalized background events from $J/\psi$ decays are estimated to be $1.1\pm0.8$ ($25.7\pm6.4$) for data sample I (II). In total, $6.9\pm1.9$ ($63.6\pm13.2$) background events are expected for data sample I (II).

The systematic uncertainties mainly arise from uncertainties in the total number of $J/\psi$ decays, the quoted intermediate BFs, the background estimation, and efficiencies associated with signal modeling, particle identification (PID), tracking of charged particles, photon detection, $\pi^0$ reconstruction, and kinematic variable requirements. The total systematic uncertainty for sample I (II) is determined to be 3.9\% (4.1\%).

Since no significant signal is observed, a maximum likelihood estimator based on the profile-likelihood approach~\cite{ref:likelihood} is used to determine the upper limit (UL) on the BF of $J/\psi\to e^\pm\tau^\mp$. The combined likelihood distribution as a function of the BF is shown in Figure~\ref{fig:likelihood} (Left). The resulting UL is $\mathcal{B}(J/\psi\to e^\pm\tau^\mp)<7.5\times10^{-8}$ at 90\% confidence level (C.L.).

\subsection{Search for CLFV decay $J/\psi\to e\mu$}

In the search for the CLFV decay $J/\psi\to e\mu$, $(8.998\pm0.040)\times10^9$ $J/\psi$ events collected in 2009, 2018, and 2019 at $\sqrt{s}=3.097~\gev$ are analyzed. Each $J/\psi$ signal candidate is reconstructed from two back-to-back oppositely charged tracks, which are further identified as one electron and one muon. 
The detection efficiency for the signal process is determined to be $(21.18\pm0.13)\%$. Based on the total $8.998\times 10^9$ $J/\psi$ events, 29 candidate events are observed in the signal region.
Two types of background events contaminate the signal region. One arises from $J/\psi$ decaying into two charged particle tracks, including $J/\psi\to \EE,~\MM,~\pi^+\pi^-,~K^+K^-,~p\bar{p}$, which is estimated to be $24.8\pm1.5$ events. The other type is continuum background from $\EE$ annihilations into pairs of charged particles, which is estimated to be $12.0\pm3.7$ events. In total, $36.8\pm4.0$ background events are expected.

The systematic uncertainties of this analysis mainly arise from efficiency estimation, tracking and PID of electrons and muons, the time-of-flight (TOF) timing difference, photon veto, and $|\Delta\theta|$ and $|\Delta\phi|$ requirements. The total systematic uncertainty is determined to be $14\%$ by adding the uncertainties from each source in quadrature.

\begin{figure}[htbp]
\centering
\includegraphics[width=0.28\textwidth]{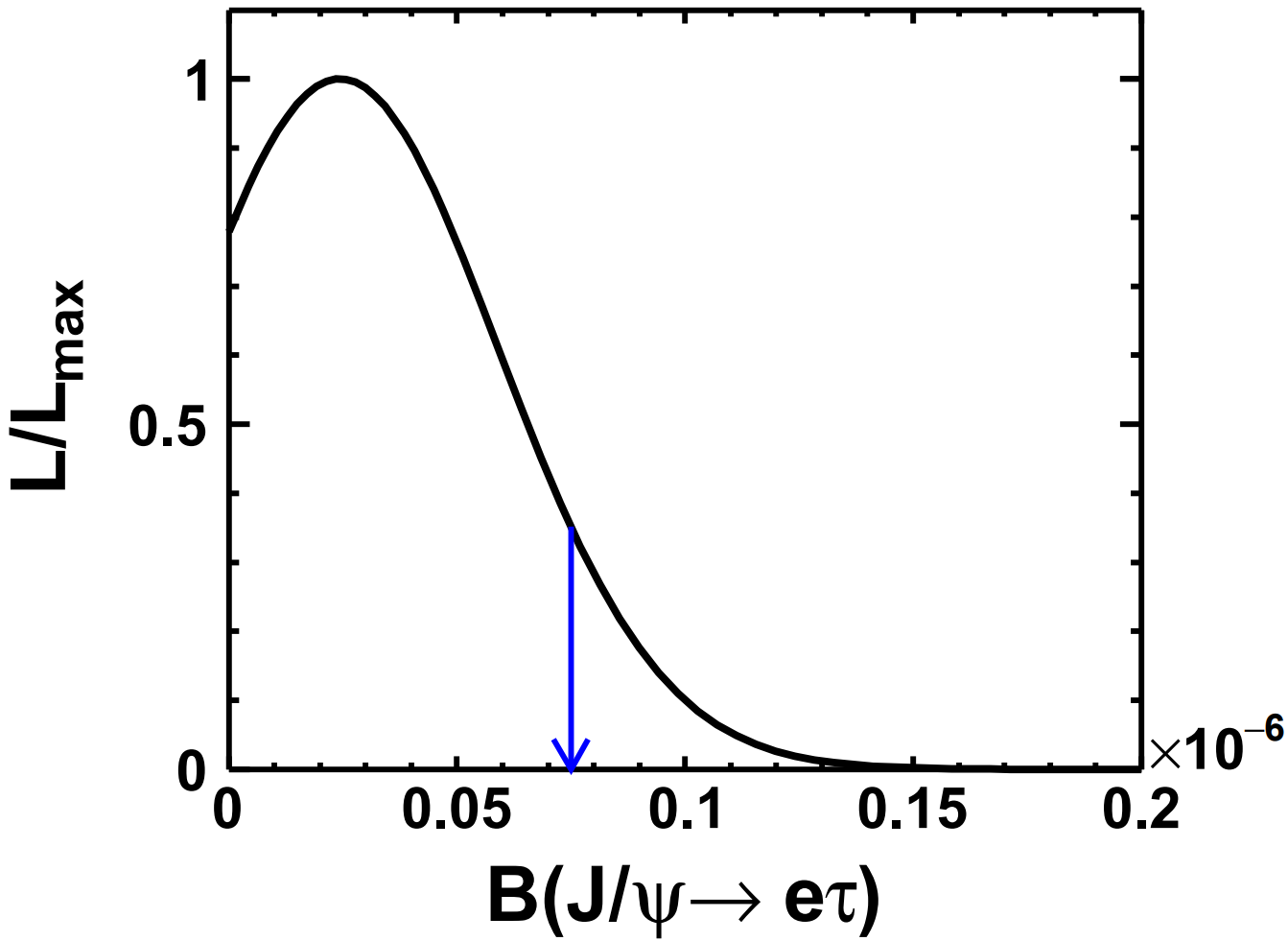}
\includegraphics[width=0.33\textwidth]{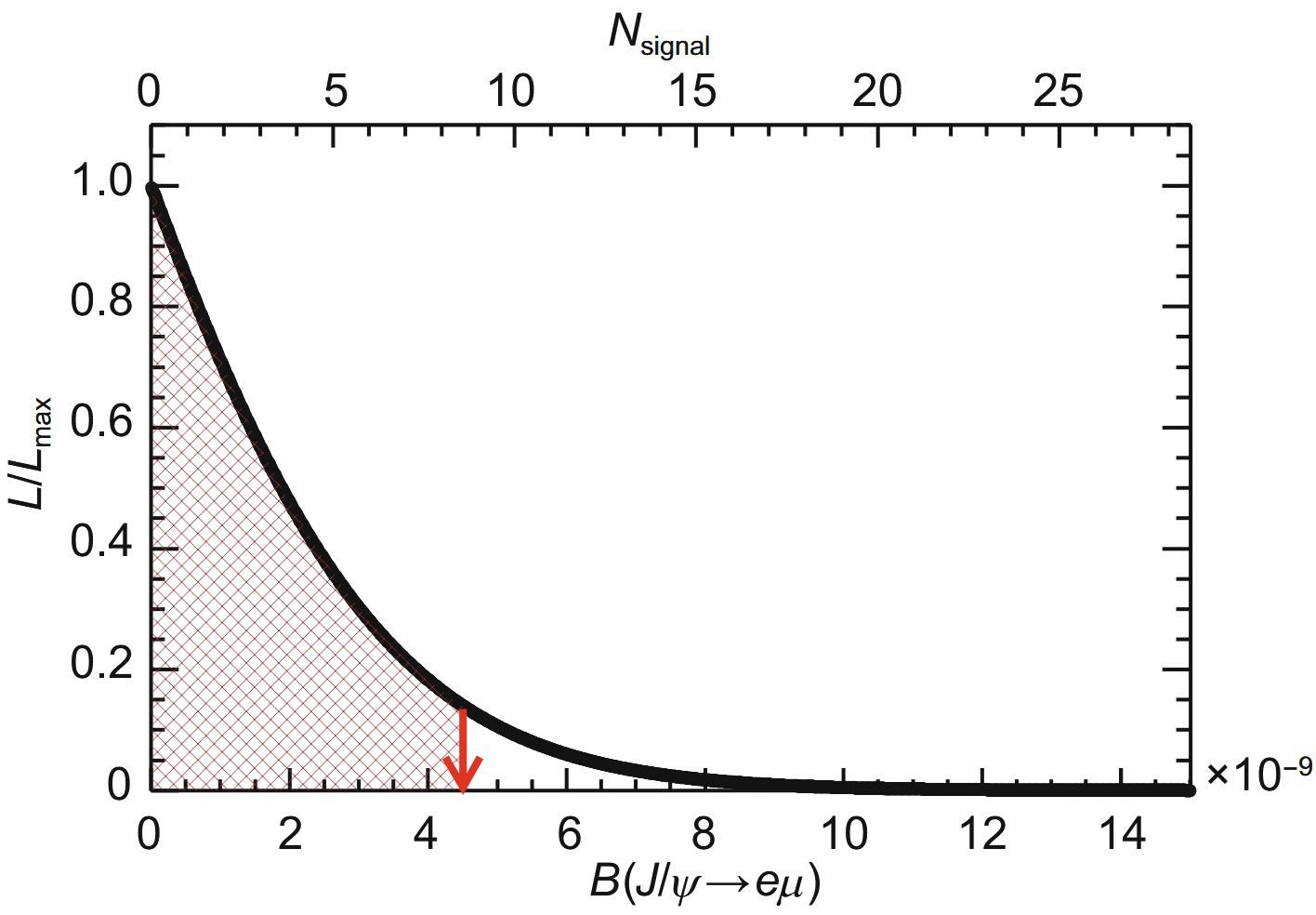}
\includegraphics[width=0.31\textwidth]{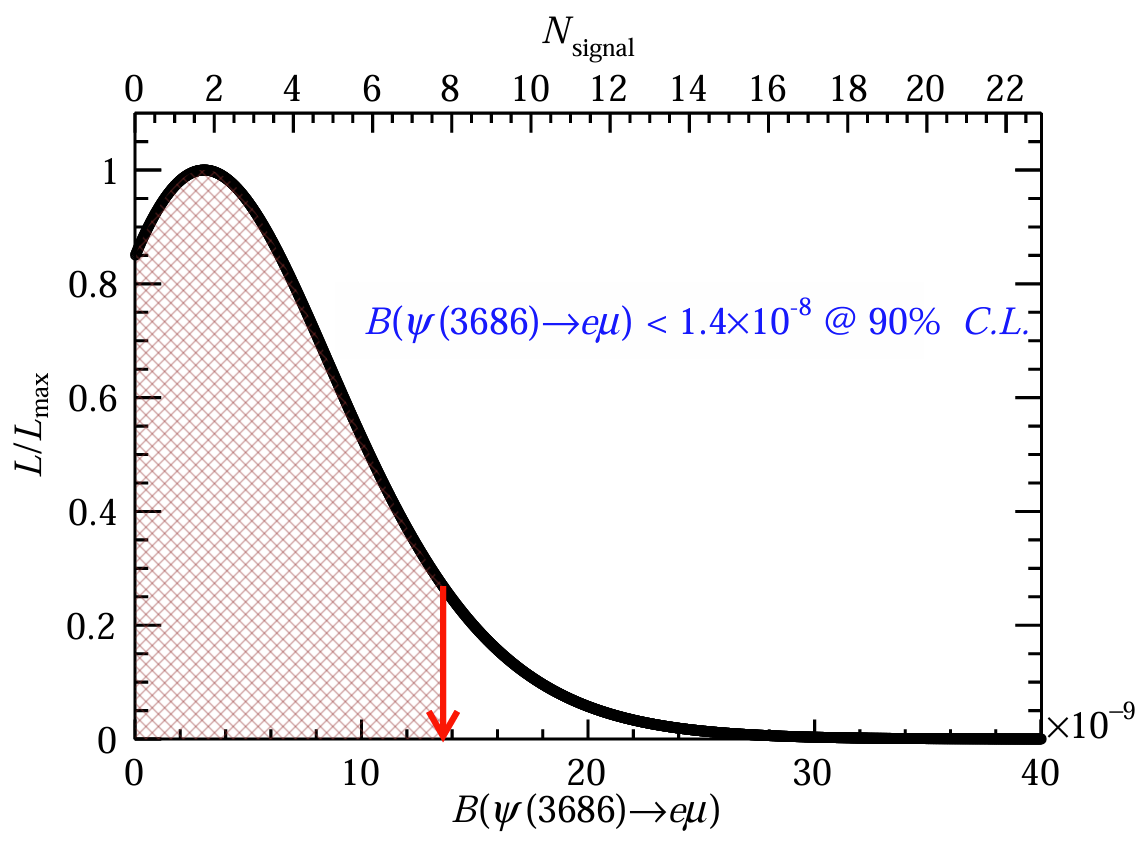}
\caption{(Left, Central, Right) Likelihood distribution as a function of the branching fraction of CLFV decays $J/\psi\to e\mu$, $J/\psi\to e\mu$, and $\psi(3686)\to e\mu$, respectively. The arrow indicates the upper limit at 90\% C.L.}
\label{fig:likelihood}
\end{figure}

In the signal region, 29 candidate events are observed, while $36.8\pm4.0$ background events are expected. Hence, no excess is observed, and an upper limit on the branching fraction $\mathcal{B}(J/\psi\to e\mu)$ is determined using the profile likelihood method. The distribution of likelihood as a function of branching fraction is shown in Figure~\ref{fig:likelihood}~(Central). The UL on the BF is found to be $\mathcal{B}(J/\psi\to e\mu)<4.5\times 10^{-9}$ at 90\% C.L. by integrating the likelihood curve in the physical region $\mathcal{B}\geq 0$.

\subsection{Search for CLFV decay $\psi(3686)\to e\mu$}

For the search for the CLFV decay $\psi(3686)\to e\mu$, the analysis method is similar to that for $J/\psi\to e\mu$. It is based on $(2.367\pm0.011)\times10^9$ $\psi(3686)$ events collected in 2009 and 2021 at $\sqrt{s}=3.686~\gev$.
The detection efficiency for the signal process is determined to be $(24.18\pm0.10)\%$ using signal MC samples. The systematic uncertainty is $11.4\%$.

Two types of background contaminate this analysis. One arises from $\psi(3686)$ decays into two charged particles, including $\psi(3686)\to \EE$, $\MM$, $\tau\tau$, $\pi^+\pi^-$, $K^+K^-$, and $p\bar{p}$. By using the event display tool, a selection highly suppressing the background is developed, which benefit this search~\cite{ref:visualbes3, Song:2025pnt}. The other is continuum background from scattering processes. The number of background events from $\psi(3686)$ decays is estimated by analyzing inclusive and exclusive MC samples and is determined to be $1.6\pm0.4$. The number of background events from continuum processes is estimated to be $4.6\pm1.4$. Eight candidate events are observed in the signal region, while $6.2\pm1.4$ background events are expected. The profile likelihood method is used to determine an upper limit on the BF $\mathcal{B}(\psi(3686)\to e^\pm\mu^\mp)$. Figure~\ref{fig:likelihood}~(Right) shows the distribution of the likelihood function. By integrating the likelihood function in the physical region $\mathcal{B}\geq0$, the UL is determined to be $\mathcal{B}(\psi(3686)\to e^{\pm}\mu^\mp)<1.4\times10^{-8}$ at 90\% C.L.

\section{Constraints on Wilson coefficients of effective field theory}

The Wilson coefficients $|C_{DL}^{\ell_{1}\ell_{2}}/\Lambda^2|$, $|C_{DR}^{\ell_{1}\ell_{2}}/\Lambda^2|$, $|C_{VL}^{q\ell_{1}\ell_{2}}/\Lambda^2|$, $|C_{VR}^{q\ell_{1}\ell_{2}}/\Lambda^2|$, $|C_{TL}^{q\ell_{1}\ell_{2}}/\Lambda^2|$, and $|C_{TR}^{q\ell_{1}\ell_{2}}/\Lambda^2|$ in effective field theory can be constrained by the ULs on the BF of charmonium CLFV decays~\cite{Hazard:2016fnc}. The constraints on Wilson coefficients are listed in Table~\ref{tab:wilson_coefficients}. 

\begin{table}[htp]
\centering
\caption{The constraints on Wilson coefficients from the results of charmonium CLFV decays.}
\vspace{0.2cm}
\label{tab:wilson_coefficients}
\begin{tabular}{cccc}
\toprule\toprule
\multirow{1}{*}{Wilson coef./$\rm GeV^{-2}$} & \multirow{1}{*}{$\ell_{1}\ell_{2}$} & $J/\psi$ & $\psi(2S)$ \\
\midrule
\multirow{2}{*}{$\abs{C_{DL/DR}^{\ell_1\ell_2}/\Lambda^2}$} 
& $e\mu$ 	& $1.8\times10^{-4}$	& $7.3\times10^{-4}$ 	\\
& $e\tau$ 	& $5.0\times10^{-5}$	& --- 					\\
\midrule
\multirow{2}{*}{$\abs{C_{VL/VR}^{q\ell_1\ell_2}/\Lambda^2}$}
& $e\mu$ 		& $1.8\times10^{-6}$ 	& $6.0\times10^{-6}$ 	\\
& $e\tau$ 	    & $1.1\times10^{-5}$ 	& ---					\\
\midrule
\multirow{2}{*}{$\abs{C_{TL/TR}^{q\ell_1\ell_2}/\Lambda^2}$} 
& $e\mu$ 		& $0.80$ 				& $2.7$ 				\\
& $e\tau$ 	    & $0.23$ 				& ---					\\
\bottomrule\bottomrule
\end{tabular}
\end{table}

\section{Summary}

In summary, we have presented searches for charged lepton flavor violation in charmonium decays, utilizing world-leading datasets collected by the BESIII experiment. ULs on the BF of CLFV decays are determined to be $\mathcal{B}(J/\psi\to e\tau)<7.5\times10^{-8}$, $\mathcal{B}(J/\psi\to e\mu)<4.5\times 10^{-9}$, and $\mathcal{B}(\psi(3686)\to e\mu)<1.4\times10^{-8}$, all at 90\% C.L. These results provide constraints on Wilson coefficients in effective field theory and probe new physics at high energy scales. Further CLFV searches at BESIII are ongoing.

\section*{Acknowledgements}

This work is supported by the National Key R\&D Program of China under Contracts Nos. 2023YFA1606000; National Natural Science Foundation of China (Grant No. 12175321, U1932101, 11975021, 11675275).

\end{document}